\documentclass[aps,preprint,groupedaddress]{revtex4}
\usepackage{bm}
\usepackage{graphicx}

\begin{document}
\bibliographystyle{apsrev}

\title{Structure of the Coulomb and unitarity corrections to the cross
 section of $e^+e^-$ pair production
in ultra-relativistic nuclear collisions}
\author{R.N. Lee}
\email[Email:]{R.N.Lee@inp.nsk.su}
\author{A.I. Milstein}
\email[Email:]{A.I.Milstein@inp.nsk.su} \affiliation{Budker
Institute of Nuclear Physics, 630090 Novosibirsk, Russia}
\author{V.G. Serbo}
\email[Email:]{serbo@math.nsc.ru} \affiliation{ Novosibirsk State
University, 630090 Novosibirsk, Russia}

\date{\today}
\begin{abstract}
We analyze the structure of the Coulomb and unitarity corrections
to the single $e^+e^-$ pair production as well as the cross
section $\sigma_n$ for the multiple pair production in collision
of ultra-relativistic nuclei. In the external field approximation
we consider the probability of one pair production at fixed impact
parameter $\rho$ between colliding nuclei. We obtain the
analytical result for this probability at large $\rho$ as compared
to the electron Compton wavelength. The energy dependence of this
probability as well as that of $\sigma_n$ differ essentially from
widely cited but incorrect results. We estimate also the unitary
corrections to the total cross section of the process.
\end{abstract}
 \maketitle

Recently the process of $e^+e^-$ pair production in
ultra-relativistic heavy-ion collisions was discussed in numerous
papers. This is connected with the beginning of operation of
Relativistic Heavy Ion  Collider (RHIC) with the Loretz factor
$\gamma=108$ and charge number of nuclei $Z=79$. New collider LHC
is scheduled to be in operation in the nearest future, with
$\gamma=3000$ and $Z=82$. The cross section of one pair production
in the Born approximation was obtained many years ago
\cite{Landau,Racah} and reads
\begin{equation}\label{Landau}
\sigma_{\rm {Born}}= \frac{28}{27\pi}\,\frac{\zeta}{m^2}\left[
L^3-2.198\,L^2+3.821\,L-1.632\right] \, ,
\end{equation}
where
\begin{equation}
\zeta=(Z_A\alpha)^2(Z_B\alpha)^2\, ,\quad L=\ln(\gamma_A\gamma_B)
\, .
\end{equation}
Here $\alpha$ is the fine-structure constant , $m$ is the electron
mass, $Z_{A,B}$ are the charge numbers of the nuclei $A$ and $B$.
The nuclei $A$ and $B$ are assumed to move in the positive and
negative directions of the $z$ axis, respectively, and have the
Lorentz factors $\gamma_{A,B}$. For the sake of simplicity we
consider the process in the frame where both nuclei have the same
Lorentz factor $\gamma_A=\gamma_B=\gamma$. In the present paper we
consider the case $L \gg 1$. Since  cross section (\ref{Landau})
is huge, a pair production can be a serious background for many
experiments. Besides, it is also important for the problem of beam
lifetime and luminosity of colliders (see review \cite{BB88}). It
means that the various corrections to the Born cross section, as
well as the cross section $\sigma_n$ for $n$-pair production, are
of great importance. At present, there is a lot of controversy in
papers devoted to this subject (the corresponding references and
critical remarks can be found in \cite{Ser,LeeM1,LeeM2}). Although
some of the corrections were obtained correctly in some special
cases, the whole consistent picture of pair production is absent.
In the present paper we intend to elucidate some points in this
problem.

For $\gamma\gg 1$ it is possible to treat  the nuclei as sources
of the external field, and calculate the probability of $n$-pair
production $P_n(\rho)$ in collision of two nuclei at a fixed
impact parameter $\rho$. The corresponding cross section
$\sigma_n$ is obtained by the integration over the impact
parameter:
\begin{equation}
\sigma_n=\int d^2{\rho}P_n(\rho)\, .
\end{equation}
The quantity which is important in further consideration is an
average number of the produced pairs at a given $\rho$:
\begin{equation}
W(\rho)= \sum_{n=1}^{\infty}n P_n(\rho)\, .
\end{equation}
The closed expression for  $W(\rho)$ was obtained in
\cite{SW,McL,Gre} :
\begin{eqnarray}\label{main}
&& W(\rho)= \int\frac{m^2d^3p\,d^3q}{(2\pi)^6\epsilon_p\epsilon_q}
\left|\int \frac{d^2k}{(2\pi)^2} \exp[{\rm i}\bm{k}\bm\rho]{\cal
M}\, {\cal F}_A({k}')\,{\cal
F}_B({k})\right|^2\,, \\
{\cal M}&=&\overline{u}(p)\left[\frac{\bm \alpha (\bm{k}-\bm
p_\perp) + \gamma_0 m}{-p_+ q_- - (\bm{k}-\bm p_\perp)^2 -m^2 }
\gamma_- +\frac{-\bm \alpha (\bm{k}-{\bm q}_\perp) + \gamma_0
m}{-p_- q_+ - (\bm{k}-{\bm q}_\perp)^2 -m^2}
\gamma_+\right]u(-q)\, .\nonumber
\end{eqnarray}
Here $\bm p$ and $\epsilon_p$ ($\bm q$ and $\epsilon_q$) are the
momentum and energy of the electron (positron), $u(p)$ and $u(-q)$
are positive- and negative-energy Dirac spinors, $\bm
\alpha=\gamma^0\bm \gamma$, $\gamma_\pm=\gamma^0\pm\gamma^z$,
$\gamma^\mu$ are the Dirac matrices, $p_\pm=\epsilon_p\pm p^z$,
$q_\pm=\epsilon_q\pm q^z$, $\bm{k}$ is a two-dimensional vector
lying in the $xy$ plane, $\bm{k}'=\bm q_\perp+\bm p_\perp-\bm{k}$,
and the function ${\cal F}(k)$ is proportional to the electron
eikonal scattering amplitude in the Coulomb field.

The function $W(\rho)$ defines the  cross section
\begin{equation}\label{sigmaT}
\sigma_T= \int d^2{\rho} W(\rho)= \sum_{n=1}^{+\infty} n\sigma_n\,
,
\end{equation}
which is called "the inclusive cross section" in \cite{McL1}. Let
us stress that the usual definition of the inclusive cross section
is different:
\begin{equation}\label{sigmaT1}
\sigma_{\rm{incl}}= \sum_{n=1}^{\infty} \sigma_n \, .
\end{equation}
To obtain $\sigma_T$ it is necessary to perform the regularization
of the expression for $W(\rho)$. One of the possible correct
regularizations is given in \cite{LeeM2} and reads
\begin{eqnarray}
\label{impact2}
 {\cal F}_{A,B}(k)&=&2\pi\int d\rho\rho J_0(\rho k)
\left\{\exp[2{\rm i}Z_{A,B}\alpha K_0(\rho a_\pm)]-1\right\} \, ,
\quad a_\pm=(p_\pm+q_\pm)/(2\gamma)\, ,
\end{eqnarray}
where $J_0$ is
the Bessel function and $K_0$ is the modified Bessel function of
the third kind. The Born cross section can be obtained  by the
replacement
\begin{eqnarray}\label{BornPropagators}
{\cal F}_{A,B}( k)\to{\cal F}^0_{A,B}(k) = \frac{4{\rm i}\pi
Z_{A,B}\alpha}{k^2+a_\pm^2}\,,
\end{eqnarray}
where ${\cal F}^0(k)$ is the first term in the expansion of ${\cal
F}(k)$ in $Z\alpha$. After the regularization the cross section
$\sigma_T$ can be presented in the form:
\begin{equation}\label{sigmaccc}
\sigma_T=\sigma_{\rm{Born}}+\sigma_T^C+\sigma_T^{CC} \, ,
\end{equation}
where $\sigma_T^C$ is the Coulomb corrections with respect to one
of the nuclei, and $\sigma_T^{CC}$ is the Coulomb corrections with
respect to both nuclei. In the main logarithmic approximation they
were obtained in \cite{LeeM1,LeeM2}
\begin{eqnarray}\label{sigmaccc1}
\sigma_T^C&=&-\frac{28}{9\pi}\frac{\zeta}{m^2}\,L^2\,[f(Z_A\alpha)+f(Z_B\alpha)]\,
,\nonumber\\
 \sigma_T^{CC}&=&\frac{56}{9\pi}\frac{\zeta}{m^2}\,L\,f(Z_A\alpha)f(Z_B\alpha)\,
 ,
\end{eqnarray}
where
\begin{equation}
f(x)=x^2\sum_{n=1}^{\infty}\frac{1}{n(n^2+x^2)} \, .
\end{equation}
The expression for $\sigma_T^C$ coincides with that
obtained in \cite{Ser} by means of
Weizs\"acker-Williams approximation. The accuracy of
the expression (\ref{sigmaccc}), (\ref{sigmaccc1}) is
determined by the relative order of the omitted terms
$\sim(Z_{A,B}\alpha)^2/L^2$. This accuracy is better
than 0.4\% for the RHIC and LHC colliders.

In a set of publications \cite{Baur90b,Roades91,BGS92,HTB95a} it
was argued that the factorization of the multiple pair production
probability is valid with a good accuracy
 \begin{equation}
\label{Pn}
P_n(\rho)=\frac{W^n(\rho)}{n!}\mbox{e}^{-W(\rho)}\,.
\end{equation}
The factor $\exp (-W)$ is nothing but the vacuum-to-vacuum
transition probability $P_0=1-\sum_{n=1}^{\infty} P_n$. Strictly
speaking, such a factorization does not take place due to the
interference between the diagrams corresponding to the permutation
of electron (or positron) lines (see, e.g., \cite{McL1}).
Nevertheless, one can show that this interference gives the
contribution which contains at least one power of $L$ less than
that of the amplitude squared.  Therefore, in the leading
logarithmic approximation we can use the expression (\ref{Pn}).

Let us represent the cross section $\sigma_1$ of one pair
production as follows
\begin{equation}\label{sigma1W}
\sigma_1=\sigma_T+\sigma_{\rm unit}=\int d^2{\rho} W(\rho) - \int
d^2{\rho} W(\rho) \left(1-\mbox{e}^{-W(\rho)}\right)
\end{equation}
Thus, the difference between $\sigma_1$ and $\sigma_T$ is due to
the unitarity correction $\sigma_{\rm unit}$.  The main
contribution to the first term ($\sigma_T$) comes from $\rho\gg
1/m$. As for the second term, $\sigma_{\rm unit}$, the main
contribution to it comes from $\rho\sim 1/m$. If $Z_{A,B}\alpha
\sim 1$, then  the exact $W(\rho)$ at $\rho\sim 1/m$ differs
essentially from the Born result $W_0(\rho)$. Since $\sigma_T$ is
studied in detail, here we investigate the unitarity correction
and $\sigma_n$ for $n\ge 2$. Below we consider two interesting
cases:

(i) $Z_{A,B}\alpha\ll 1$, $\zeta\,L\ll 1$ ;

(ii) $Z_{A,B}\alpha\ll 1$, $\zeta\, L\sim 1$.

Let us consider in detail the case (i), where it is possible to
use $W_0$ (obtained in the Born approximation) instead of the
exact function $W$, expand the exponent $\exp(-W)$ in the
unitarity correction and omit it in $\sigma_n$ for $n\ge 2$. For a
few particular values of $\gamma$ the function $W_0(\rho)$ was
calculated numerically in \cite{HTB95b, Gucl95} around $m\rho\sim
1$ . In \cite{BB88}, for $m\rho\gg 1$, this function was
approximated by a simple expression
 \begin{equation}\label{WBB}
\tilde W_0(\rho) = {14\over 9\pi^2} \, {\zeta\over (m\rho)^2}\;
\left(\ln{0.681 \gamma^2\over m\rho}\right)^2\,,\;\; 1 \ll m\rho
\ll \gamma^2 \, .
\end{equation}
This expression looks very convenient for fast estimates of
various quantities. That is why Eq. (\ref{WBB}) is widely cited
and used in many papers (see, e.g.,
Refs.~\cite{Baur90a,Baur90b,Roades91,HTB95b,AHTB97,BHT98,GLUES99,
Guclu2000}). Now we show that Eq.(\ref{WBB}) is incorrect. We find
out that there are two scales in dependence of $W_0(\rho)$ on
$\rho$: in the region of relatively small impact parameters, $1\ll
m\rho \le \gamma$, we  obtain
 \begin{equation}\label{WB1}
W_0(\rho) = {28\over 9\pi^2}\, {\zeta\over (m\rho)^2}\;
\left[2\ln{ \gamma^2}-3\ln{( m\rho)}\right]\, \ln{(m\rho)} \, ,
\end{equation}
 while in the region of relatively large impact parameters,
 $\gamma \le m\rho \ll \gamma^2$, we have
\begin{equation}\label{WB2}
W_0(\rho) = {28\over 9\pi^2}\, {\zeta\over (m\rho)^2}\;
\left(\ln{\gamma^2\over m\rho}\right)^2\,.
\end{equation}
Note that the function $W_0(\rho)$ given by Eqs. (\ref{WB1}) and
(\ref{WB2}) is the continuous function at $m\rho= \gamma$ together
with its first derivative. Certainly, the integration of
$W_0(\rho)$ from Eqs. (\ref{WB1}) and (\ref{WB2}) over $\bm \rho$
gives the main term ($\propto L^3$) in Eq.(\ref{Landau}).

 The most important distinction between
Eq.({\ref{WBB}) and our result is a quite different dependence of
$W_0(\rho)$ on $\gamma$ for $1\ll m\rho \ll \gamma$. It follows
from Eq.(\ref{WB1}) that $W_0(\rho)\propto L$ at $m\rho \sim 1$
while Eq.(\ref{WBB}) gives $\tilde W_0(\rho)\propto L^2$ at $m\rho
\sim 1$. Since $\sigma_{\rm unit}$ and $\sigma_n$ (for $n\ge 2$)
are determined by the region of integration $m\rho\sim 1$, the
prediction obtained with the help of Eq. (\ref{WBB}) leads to
incorrect dependence of these quantities on $\gamma$ (see below).

Let us consider the derivation of Eqs. (\ref{WB1}) and
(\ref{WB2}). The main contribution to the integrals in
(\ref{main}) at $\rho\gg 1/m$ comes from the region of integration
$$
|\bm{k}|,\,|\bm p_\perp+\bm q_\perp|\ll m\,,\qquad|p_z|,\,|q_z|\ll
m\gamma\,,\qquad |\bm p_\perp-\bm q_\perp|\sim m\,.
$$
Passing to the variables $\bm P=\bm p_\perp+\bm q_\perp$,
$\bm{r}=(\bm p_\perp-\bm q_\perp)/2$, $E=p_z+q_z$, and
$x=(p_z-q_z)/(p_z+q_z)$, we  take the integrals over $\bm r$ and
$x$ (see \cite{LeeM2}). Then, we obtain:
\begin{eqnarray}
W_0(\rho)&=&\frac{112}{9\pi}\frac{\zeta}{m^2}\int_m^{m\gamma}\frac{dE}{E}
\int d^2 P\, L^{ij}L^{ij\,*}\, ,\nonumber\\
L^{ij}&=&\int \frac{d^2k\, k^i\,(P-k)^j\exp({\rm i}\bm{k\rho})}
{(2\pi)^2\,[{\bm k}^2+E^2/\gamma^2][(\bm P-\bm
k)^2+m^4/(E^2\gamma^2)]}\, .
\end{eqnarray}
After the integration over $\bm k$ we come to
\begin{eqnarray}
W_0(\rho)&=&\frac{56}{9\pi^2}\frac{\zeta}{m^2}\int_m^{m\gamma}\frac{dE}{E}
\left[\int_{1/\rho}^{m} \frac{dP}{P}\vartheta(\gamma/E-\rho)+
\int_{\max\{1/\rho,E/\gamma\}}^{m} \frac{dP}{P}\vartheta(\gamma
E/m^2-\rho)\right]\, ,
\end{eqnarray}
where $\vartheta(x)$ is the step function. Straightforward
integration leads to Eqs. (\ref{WB1}), (\ref{WB2}). From our
derivation it is clear that  two large logarithms at $m\rho\gg 1$
come from the integration over $E$ and $P$, while at $m\rho \sim
1$ the logarithm from the integration over $P$ is absent and the
only logarithm $L$ arises from the integration over $E$.

The result (\ref{WB1}), (\ref{WB2}) can also be obtained within
the standard Weizs\"acker-Williams approximation. In this
approximation the colliding nuclei with impact parameters $\rho_1$
and $\rho_2$ emit $dn_1$ and $dn_2$ equivalent photons with
energies $\omega_1$ and $\omega_2$:
\begin{equation}
 dn_i = {Z_i^2\alpha\over \pi^2}\, {d\omega_i \over \omega_i}\, {d^2
\rho_i\over \rho_i^2}\,, \;\; i=1,\,2\,, \quad\omega_i \ll
m\gamma\,, \;\; \frac{1}{m}\ll \rho_i \ll
\frac{\gamma}{\omega_i}\,.
\end{equation}
 Then these photons collide and produce a pair with the
invariant mass squared $s=4\omega_1\omega_2$. Therefore, the
function $W_0(\rho)$ is
 \begin{equation}
W_0(\rho) = \int dn_1 dn_2 \delta(\bm \rho_1 -\bm \rho_2-
 \bm \rho)\,\sigma_{\gamma\gamma}
(s)\,
\end{equation}
 where $\sigma_{\gamma \gamma}$ is the cross section of the process
$\gamma \gamma \to e^+e^-$. Then we perform the integration over
$\omega_2$ using the relation
\begin{equation}
\int^\infty_{4m^2}\,{ds\over s}\,\sigma_{\gamma \gamma} (s) =
{14\pi\over 9}\, {\alpha^2\over m^2}\,.
 \end{equation}
The main contribution to this integral is given by the region near
the lower integration limit $s= 4m^2$. That it is why we can
extend the upper limit up to infinity and substitute $\omega_2 =
m^2/\omega_1$ in the step function. After that we have
\begin{eqnarray}
W_0(\rho)=\frac{14}{9\pi^3}\,\frac{\zeta}{m^2}\,
\int_{m/\gamma}^{m\gamma}\frac{d\omega_1}{\omega_1}
\int\frac{d^2\rho_1}{\rho_1^2(\bm \rho-\bm\rho_1)^2}\,
\vartheta\left(\frac{\gamma}{\omega_1}-\rho_1\right)\,
\vartheta\left(\frac{\gamma\omega_1}{m^2}-|\bm\rho-\bm
\rho_1|\right)\, .
\end{eqnarray}
The main contribution to this integral is given by two regions:
$1/m\ll \rho_1\ll \rho$ and $1/m\ll |\bm\rho-\bm \rho_1|\ll \rho$.
Then we have
\begin{eqnarray}
W_0(\rho)&=&\frac{28}{9\pi^2}\,\frac{\zeta}{(m\rho)^2}\,
\int_{m/\gamma}^{m\gamma}\frac{d\omega_1}{\omega_1}
\int_{1/m}^\rho\frac{d\rho_1}{\rho_1}\,\nonumber\\ &&\times\left[
\vartheta(\gamma/\omega_1-\rho_1)\,
\vartheta(\gamma\omega_1/m^2-\rho)+\vartheta(\gamma/\omega_1-\rho)\,
\vartheta(\gamma\omega_1/m^2-\rho_1)\right]\, .
\end{eqnarray}
The further integration leads again to (\ref{WB1}), (\ref{WB2}).

As we argued above, the function $W_0(\rho)$ at $m\rho\ll \gamma$
and $L\gg 1$ has the form
\begin{equation}\label{Eq:Fdef}
W_0(\rho)=\zeta\,L\, F(m\rho)\,
\end{equation}
with the universal function $F(x)$ independent of $Z_{A,B}$ and
$\gamma$.
\begin{figure}[h]
  \centering
  \includegraphics[height=170pt,keepaspectratio=true]{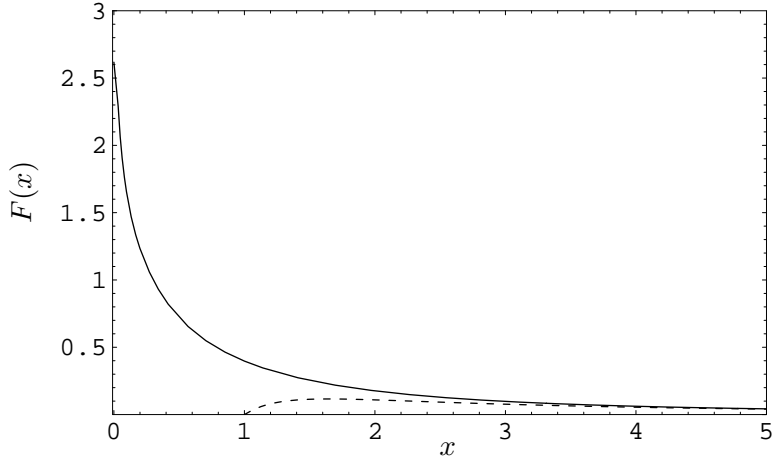}
 \begin{picture}(0,0)(0,0)
 \put(-136,-3){ $x$}
 \put(-295,85){\rotatebox{90}{$F(x)$}}
 \end{picture}
  \caption{The function $F(x)$ from Eq. (\ref{Eq:Fdef}) vs.
  $x=m\rho$ (solid curve) and its asymptotic form
  $F_{\rm asymp}(x)$, Eq. (\ref{Eq:Fasymp})(dashed curve).}
  \label{Fig:F}
\end{figure}

We extract this function (see Fig. \ref{Fig:F}) from the numerical
results in \cite{HTB95b} for the case $\gamma=3400$. It follows
from (\ref{WB1}) that the asymptotic form of the function $F(x)$
at $x\gg 1$ reads
\begin{equation}\label{Eq:Fasymp}
F_{\rm asymp} (x)=\frac{56\ln x}{9\pi^2 x^2}\,.
\end{equation}

Using this function we obtain in the region (i) the unitarity
correction to the one-pair production cross section
\begin{equation}\label{unit1}
\sigma_{\rm unit}=-2C_2\,\frac{\zeta^2\,L^2}{m^2} \,
\end{equation}
and the cross sections for the n-pair production ($n\ge 2$)
\begin{eqnarray}\label{Cn}
\sigma_{n}&=&C_n \frac{\zeta^n L^n}{m^2}\,
,\\%\nonumber\\
 C_n&=&\frac{2\pi}{n!}\int_0^{\infty}\,F^n(x)\,x dx
 \, ,\\%\nonumber\\
 C_2&=&1.33\, ,\quad C_3=0.264\, ,\quad C_4=0.066\,
,\quad C_5=0.0176\, .
\end{eqnarray}
Our result $\sigma_n\propto L^n$ for $n\ge 2$ differs considerable
from the incorrect results $\sigma_n\propto L^{2n}$ of
\cite{Baur90b} and $\sigma_n\propto L^{3n}$ of \cite{Guclu2000}.

Let us pass to the consideration of the case (ii). If $\zeta\,
L\sim 1$, but $\zeta\, (Z_{A,B}\alpha)^2\,L\ll 1$, then we can
neglect the Coulomb corrections in $W(\rho)$ but should keep the
exponent in Eq.(\ref{Pn}). It gives the result similar to Eq.
(\ref{Cn}) for $\sigma_n$ with the replacement
\begin{equation}
C_n\to \tilde
C_n(\gamma,Z_{A,B})=\frac{2\pi}{n!}\int_0^{\infty}\,F^n(x)
\exp[-\zeta\,L\, F(x)] \,x dx \, .
\end{equation}
For unitarity correction we have
\begin{equation}\label{unit2}
\sigma_{\rm unit}=-{2\pi}\frac{\zeta}{m^2}\,
L\,\int_0^{\infty}\,F(x)\left\{1- \exp[-\zeta\,L\, F(x)]\right\}
\,x dx \, .
\end{equation}

If $\zeta\,(Z_{A,B}\alpha)^2 L\sim 1$, then we should use in Eq.
(\ref{Pn})  the function  $W(\rho)$ calculated exactly with
respect to the parameters $Z_{A,B}\alpha$. Note that $W(\rho)$,
for $L\gg 1$ and $m\rho\sim 1$, has the form similar to
Eq.(\ref{Eq:Fdef})
\begin{equation}\label{Eq:Ftildedef}
W(\rho)=\zeta\,L\, \tilde{F}(m\rho,Z_A\alpha,Z_B\alpha)\,.
\end{equation}
The function $\tilde{F}(m\rho,Z_A\alpha,Z_B\alpha)\to F(m\rho)$
for $Z_{A,B}\alpha\to 0$, but the difference $\tilde{F}-F$ can be
neglected in the exponential factor $\exp (-W)$ only if this
difference is small as compared to $1/(\zeta L)$ rather than to
$F$.

\begin{figure}[h]
  \centering
  \includegraphics[height=170pt,keepaspectratio=true]{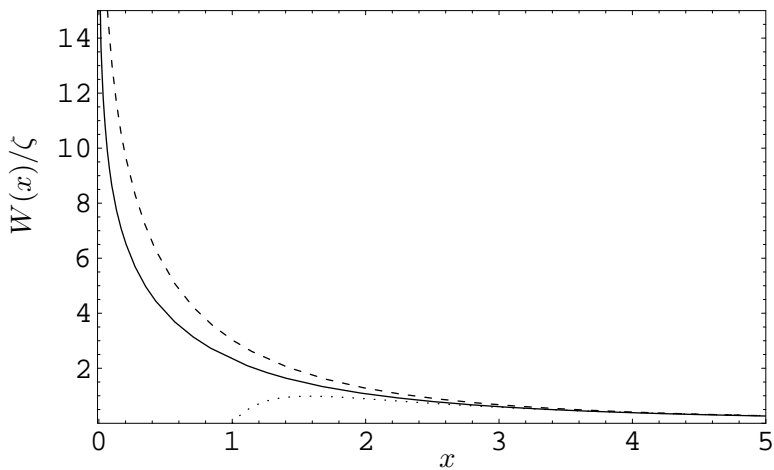}
 \begin{picture}(0,0)(0,0)
 \put(-136,-3){ $x$}
 \put(-295,85){\rotatebox{90}{$W(x)/\zeta$}}
 \end{picture}
  \caption{The exact function $W(x)$ (solid curve),
  Born approximation $W_0(x)$ (dashed curve), and the asymptotic
  form of $W_0(x)$ from (\ref{WB1}), (\ref{WB2}) (dotted curve)
  in units of $\zeta$.}
  \label{Fig:W}
\end{figure}
The function $W(\rho)$ was calculated numerically in \cite{HTB99}
for the particular case $\gamma=100$, $Z=79$. In Fig. \ref{Fig:W}
it is shown together with $W_0(x)$, calculated also in
\cite{HTB99}, and our asymptotics (\ref{WB1}), (\ref{WB2}). There
is a good agreement of our analytical results with the exact
numerical one  already  at $m\rho>2$. It is also seen a noticeable
difference between $W(\rho)$ and $W_0(\rho)$ in the region
$m\rho\sim 1$, that is essential in calculations of the unitarity
correction and $\sigma_n$ for $n\ge 2$. Emphasize that the Coulomb
corrections $\sigma^C_T\propto L^2$ in (\ref{sigmaccc1}) arise due
to the difference between $W(\rho)$ and $W_0(\rho)$ in the region
$m\rho\gg 1$.

Using the numerical results for $W(\rho)$ and $W_0(\rho)$ from
\cite{HTB99} we find that the exact value of $\sigma_{\rm
unit}/\sigma_{\rm Born}$ is equal to $-4.1$\% , while the result
without Coulomb effects gives $-6.4$\% . With the help of
Eqs.(\ref{unit2}) and (\ref{unit1}) we obtain $-9.3$\% and
$-12$\%, respectively. Thus, in the example considered
($\gamma=100$, $Z=79$) the account of the Coulomb effects as well
as the exponential factor is very important.

It is interesting to estimate the unitarity correction for the LHC
case ($\gamma=3000,\, Z=82$). Using the recent numerical results
of K.~Hencken (private communication) for $W(\rho)$ at
$\gamma=3400$, $Z=82$, we find that the exact value of
$\sigma_{\rm unit}/\sigma_{\rm Born}$  for this case is equal to
$-3.2$\%. This ration remains approximately the same for the LHC
case.

We are very grateful to K. Hencken for sending us the numerical
data cited above. V.G.S. would like to thank A. Baltz, F. Gelis,
L. McLerran, and A. Peshier for useful discussions during his stay
in BNL. This work was supported through Grants RFBR 00-02-17592,
01-02-16926, and St.-Petersburg E 00-3.3-146.

\end{document}